\documentclass[
    ,final            
  ]
  {aipproc}

\layoutstyle{6x9}
\usepackage{graphicx,amsmath,amssymb,bm}

\begin{document}

\title{New applications of Renormalization Group methods to nuclear matter}


\author{Kai Hebeler}{
  address={Department of Physics, The Ohio State University,
Columbus, OH 43210, USA}
}

\begin{abstract}
We give an overview of recent results for the nuclear equation of state and properties of neutron stars based 
on microscopic two- and three-nucleon interactions derived within chiral effective field theory (EFT). It is demonstrated that 
the application of Renormalization Group (RG) transformations allows efficient, controlled and simplified calculations of nuclear matter.
\end{abstract}

\maketitle

\section{The Similarity Renormalization Group}

Establishing an interparticle Hamiltonian, which is the fundamental ingredient for microscopic 
many-body calculations of atomic nuclei and nucleonic matter, is a difficult and on-going challenge 
for low-energy nuclear physics. Chiral EFT and RG methods make feasible a 
controlled description of nuclear interactions, grounded in QCD symmetries, which in turn makes possible the 
description of nuclei across the nuclear many-body landscape~\cite{Epelbaum_RevModPhys, Bogner_review}.
The two-body sector has been solved in the sense that various 
nucleon-nucleon interactions are available that reproduce low-energy scattering phase shifts to high accuracy. The 
unsettled frontier is three- and higher-body forces, which play a key role in many-nucleon systems. 

\begin{figure}[t]
  \includegraphics[width=0.95\textwidth,clip]{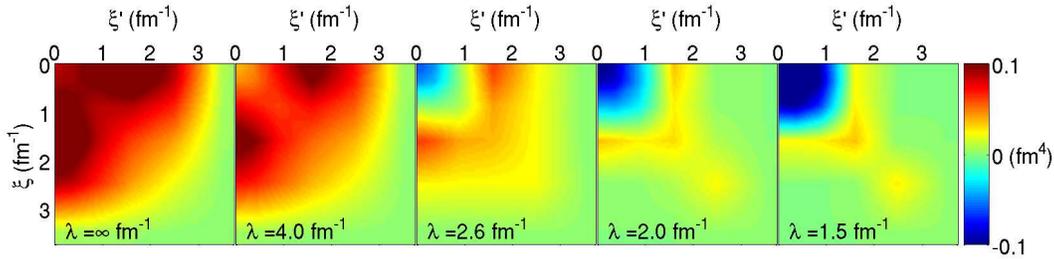}
  \caption{(from Ref.~\cite{Hebeler_3NF_evolution}) Contour plot of the RG-evolved 3N potential as a function of the hypermomenta $\xi$ and $\xi'$ for a fixed hyperangle. Evolution proceeds from left to right, with softening evidence by the suppression of off-diagonal elements.
}
\label{fig:3NF_decoupling}
\end{figure}

Nuclear structure calculations are complicated due to the coupling of low to
high momenta by nuclear interactions. A fruitful path to decoupling high-momentum
from low-momentum physics is the Similarity Renormalization Group (SRG), which is based on a 
continuous sequence of unitary transformations that suppress off-diagonal matrix elements. The decoupling 
of momenta via the SRG has been demonstrated for nuclear nucleon-nucleon (NN) interactions~\cite{Bogner_SRG_evolution} and very recently for the first 
time also for three-nucleon forces (3NF)~\cite{Hebeler_3NF_evolution}. The decoupling in 3NF is illustrated in Fig.~\ref{fig:3NF_decoupling}: 
at large resolution scales $\lambda = \infty$ the potential contains significant 
off-diagonal couplings. As we evolve to lower resolution, these couplings get successively suppressed and finally at 
$\lambda = 1.5\:\rm{fm}^{-1}$ the non-perturbative features are substantially softened as we find only non-vanishing strength at
small momenta and around the diagonal.

SRG-evolved potentials are automatically energy
independent and the same transformations renormalize  all operators, including
many-body operators, and the class of transformations can be tailored for 
effectiveness in particular problems. When evolving nuclear interactions to lower 
resolution, it is inevitable that many-body interactions
and operators are induced even if initially absent. This 
might be considered as unnatural if nuclei could be accurately calculated based on only NN
interactions, as was assumed for much of the history of nuclear structure calculations. 
However, chiral EFT reveals the natural scale and hierarchy of many-body forces, which 
dictates their inclusion in modern calculations of nuclei and nucleonic matter.

\section{Nuclear equation of state}

Progress towards controlled nuclear matter calculations has
long been hindered by the non-perturbative nature of the nuclear many-body problem
when conventional nuclear potentials are used. The novel developments described above open 
new ways to overcome these obstacles. Calculations of finite nuclei based on consistently evolved NN \emph{and} 
3N potentials are now available (see Refs.~\cite{Jurgenson_PRL_PRC}). Such fully consistent calculations of the nuclear
equation of state do not easily follow, but the framework recently presented in Ref.~\cite{Hebeler_3NF_evolution} makes them
likely in the near future.

\begin{figure}[t]
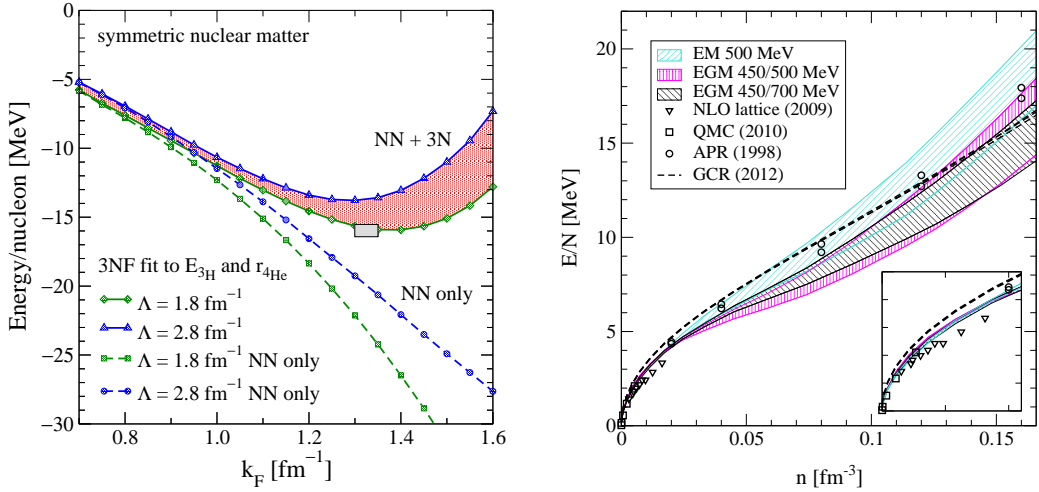

  \includegraphics[width=0.45\textwidth,clip]{429_Hebeler-f2.eps}
  \hspace{0.5cm}
  \includegraphics[width=0.43\textwidth,clip]{429_Hebeler-f3.eps}
  
  \caption{left panel: (from Ref.~\cite{Hebeler_SNM}) Nuclear matter energy as a function 
of the Fermi momentum based on NN+3NF forces compared to NN-only results for two representative NN cutoffs 
and a fixed 3N cutoff; right panel: (from Ref.\cite{Tews_N3LO}) Neutron matter energy as a function of density 
including NN, 3N and 4N forces up to N$^3$LO in comparison to other studies.}
\label{fig:EOS_SNM_PNM}
\end{figure}

As an alternative simplified strategy it is also possible to evolve only the NN interactions with RG methods 
and then fix the the short-range parameters of the 3N forces from fits to few-body systems 
at the low-momentum scale~\cite{Hebeler_SNM}. This procedure assumes that the long-range part of the 3N forces remains 
invariant under the RG transformations and that the operator structure of the chiral interaction is a sufficiently 
complete basis so that induced contributions can be absorbed to good approximation. 
As shown in the left panel of Fig.~\ref{fig:EOS_SNM_PNM}, we find realistic saturation properties within
our theoretical uncertainty bounds without adjustment of free parameters. The two pairs of curves show 
the difference between the nuclear matter results for NN-only
and NN plus 3N interactions. It is evident that saturation is driven by 3NF. 
Even for $\Lambda = 2.8\,\rm{fm}^{-1}$, which is similar to the lower 
cutoffs in chiral EFT potentials, saturation is at too high density without the 3NF.

Neutron matter provides a different perspective to nucleonic matter. Here only the long-range 
2$\pi$-exchange 3NF contribute~\cite{Hebeler_PNM}, which implies that all 
three- and four-neutron (4N) forces are predicted up to N$^3$LO. The physics of neutron matter ranges 
from the universal regime at low densities that can be probed in experiments with ultracold atoms up to 
high densities relevant for the structure of neutron stars. For these extreme conditions, controlled 
calculations with theoretical error estimates are essential. As a result of effective range effects, restricted 
phase space at finite density and weaker tensor forces between neutrons
neutron matter behaves more perturbative than symmetric nuclear matter. In the right panel 
of Fig.~\ref{fig:EOS_SNM_PNM} we present results for neutron matter based directly on chiral EFT interactions 
without RG evolution including all NN, 3N and 4N contributions up to N$^3$LO~\cite{Tews_N3LO}. We find relatively 
large attractive contributions from N$^3$LO 3N forces, which might be an indication that a chiral EFT with
explicit $\Delta$ degrees of freedom may be more efficient~\cite{Epelbaum_RevModPhys}. In contrast, contributions from 4N 
interactions appear to be very small in neutron matter.
 
\section{Astrophysical applications}

\begin{figure}[t]
  \includegraphics[width=0.45\textwidth,clip]{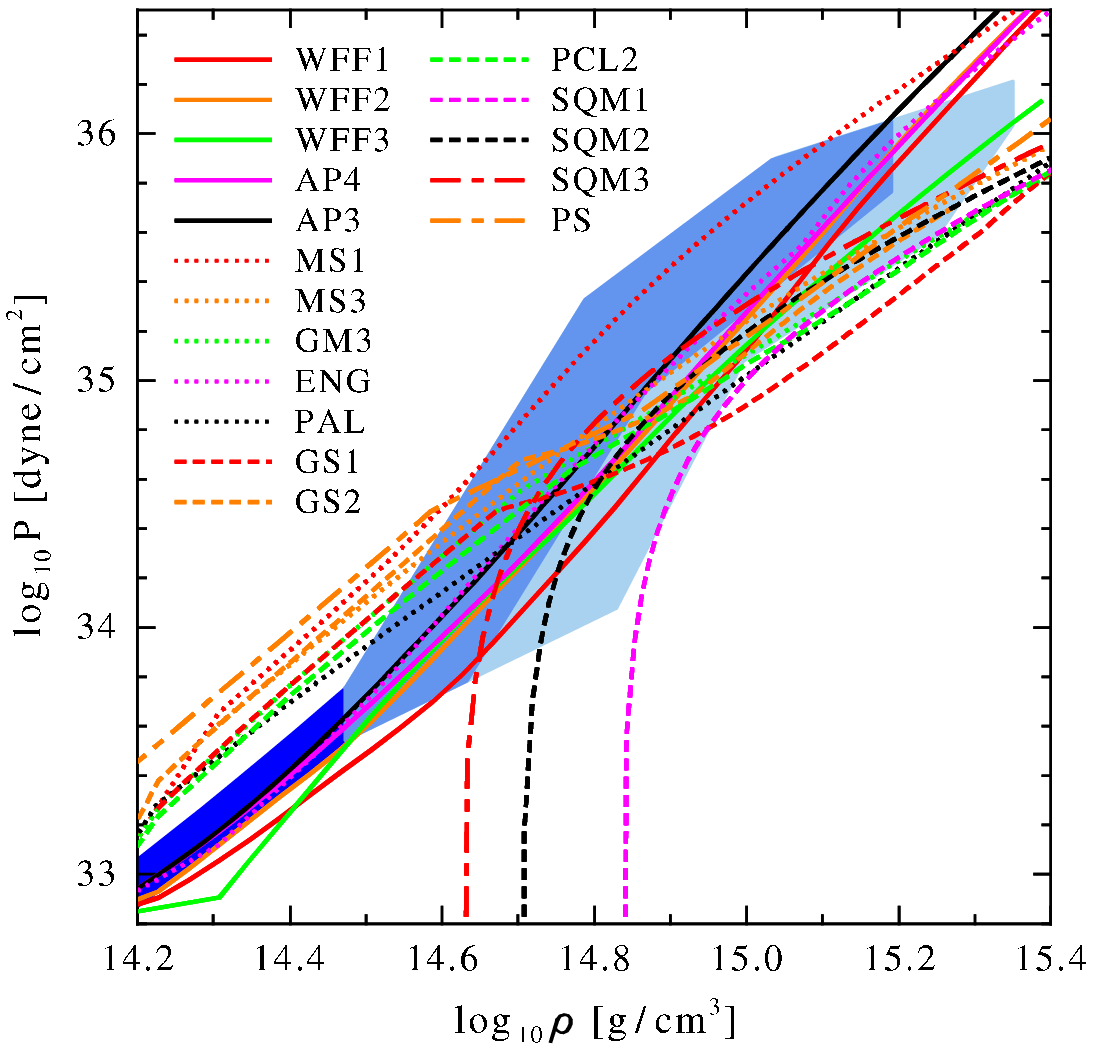}
  \hspace{0.5cm}
  \includegraphics[width=0.45\textwidth,clip]{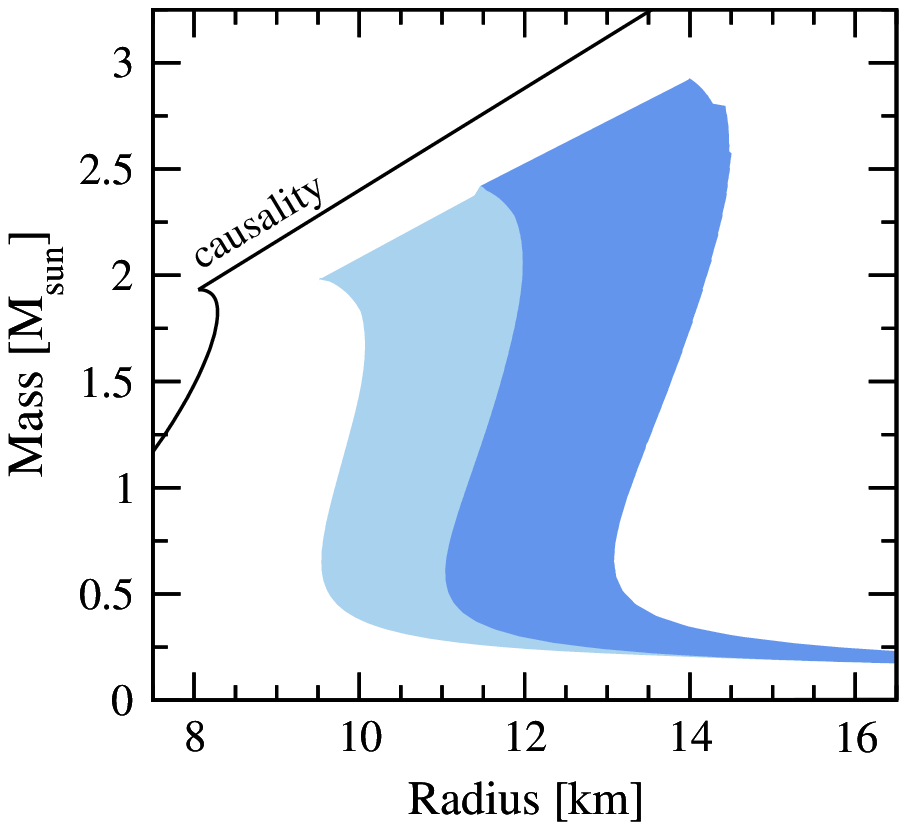}
  \caption{left panel: uncertainty of the pressure as a function of mass density 
    compared to a representative set of equations of state used in the literature (see Ref.~\cite{LP});
    right panel: corresponding uncertainty of neutron star radius and mass. For both panels, the light band corresponds 
    to the mass constraint $M_{\rm{min}} = 1.97\,M_{\odot}$ and the dark band to
    $M_{\rm{min}} = 2.4\,M_{\odot}$. See Ref.~\cite{Hebeler_NS_long} for details.}
  \label{fig:NS}
\end{figure}

Core densities inside a neutron star can reach several times nuclear 
saturation density. For this reason, we need to extend the microscopically calculated
EOSs to higher densities. To do so we use a general strategy which 
does not rely on assumptions about the nature of the constituents 
of the matter and their interactions at higher densities: 
we employ a piecewise polytopic ansatz $P(\rho) = \kappa \, \rho^{\Gamma}$~\cite{Hebeler_NS_long} 
(see also Refs.~\cite{poly, Hebeler_NS}), whereas the values of the parameters are 
limited by physics and constraints from observations. As constraints we require that (a) the 
speed of sound remains smaller than the speed of light for all densities, and (b) the EOS is 
able to support a neutron star of mass $M \ge M_{\rm{min}}$, where the value of $M_{\rm{min}}$ is given by experimental 
neutron star observations. For $M_{\rm{min}}$ we consider two cases: first, $M_{\rm{min}}= 1.97\,M_{\odot}$, which is currently the
heaviest confirmed observed neutron star mass~\citep{Demorest}, and second, a hypothetical mass $M_{\rm{min}}=2.4\,M_{\odot}$. 
We generate a very large number of EOSs for different values of parameters and retain only those which 
fullfill the causality and mass conditions. This results in uncertainty bands for the pressure and neutron star 
radii, which are shown in Fig.~\ref{fig:NS}. In the left panel we present the uncertainty bands for the equation 
of state for the two mass constraints compared to a representative set of other EOSs~(see Ref.~\cite{LP}). We find that only 
a relatively small number of EOSs are compatible with the extracted uncertainty bands for all densities. In the right panel we 
show the resulting constraints for the radius of neutron stars. For a typical neutron star of 
mass $M=1.4\,M_{\odot}$ we find a radius range $R=10.0-13.7\:\rm{km}$ for $M_{\rm{min}}= 1.97\,M_{\odot}$ and $R=11.6-13.7\:\rm{km}$ 
for $M_{\rm{min}}= 2.4\,M_{\odot}$.

\section{Correlations in nuclear systems}

Recent experimental studies of proton knock-out reactions off nuclei at high-momentum transfer have been explained 
by invoking short-range correlations (SRC) in nuclear systems~\cite{SRCScience}. Such explanations may seem at odds 
with RG evolution, which leads to many-body wave functions with highly suppressed SRC. The key is that not only the 
Hamiltonian but every other operator must also evolve. Recent advances make it possible to 
systematically evolve operators using the SRG as shown in Ref.~\cite{EricOperatorEvolv}.

\begin{figure}[t]
  \includegraphics[scale=0.35,clip]{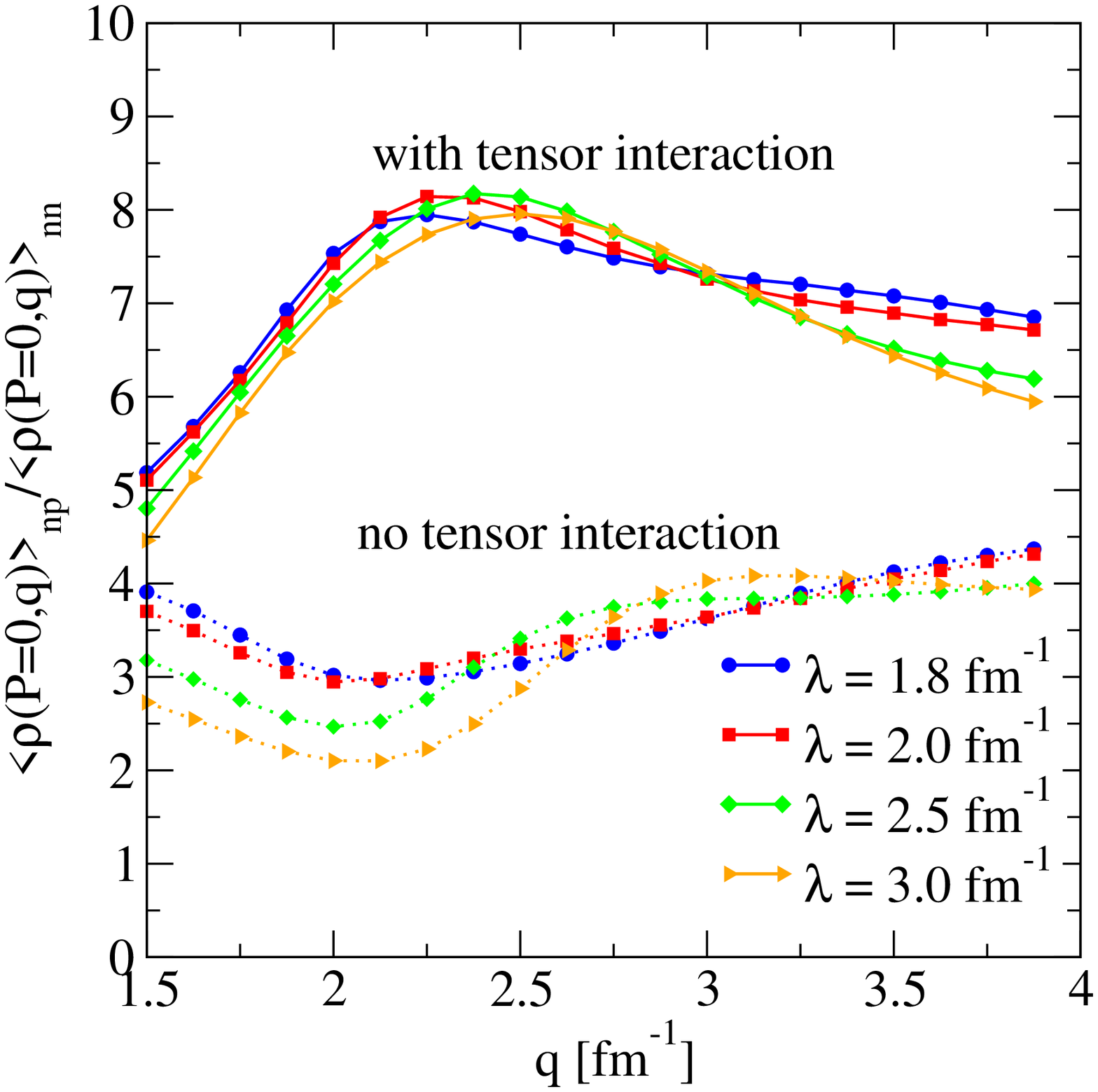}
  \hspace{0.5cm}
  \includegraphics[scale=0.35,clip]{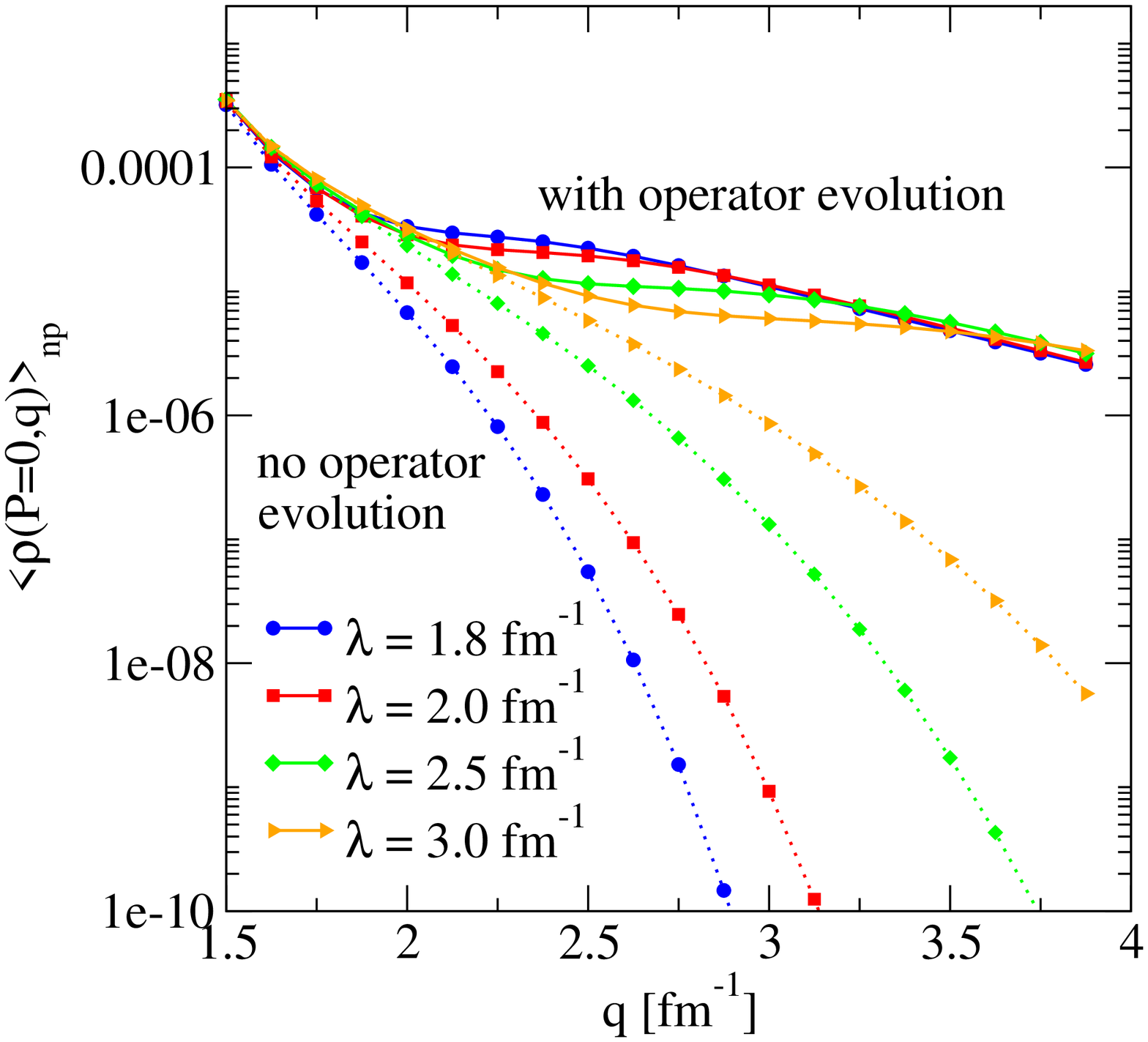}
  \caption{The pair density function in nuclear matter at saturation density as a function of relative momentum $q$ for 
    different SRG resolution scales $\lambda$. The left panel demonstrates the importance of a consistent evolution of 
    Hamiltonian and density operator, the right panel the significance of tensor forces in the initial Hamiltonian, leading to an 
    enhancement of neutron-proton pairs over neutron-neutron pairs.}
  \label{fig:SRC_NM}
\end{figure}

In Fig.~\ref{fig:SRC_NM} we show first preliminary results for the pair momentum distribution $\rho(P,q)$ in nuclear 
matter as a function of the relative momentum $q$, obtained within a perturbative expansion~\cite{SRC_NM_inprep}. Here $P$ 
denotes the center-of-mass momentum of the pair. We choose as initial 
Hamiltonian the highly non-perturbative Argonne V18 NN potential and evolve consistently density 
operator and Hamiltonian via the SRG, each truncated at the two-body level. As shown in the left panel, already at this level 
of approximation, the pair momentum distribution function is invariant to good approximation in back-to-back kinematics, in contrast
to the case without operator evolution (dotted lines). In addition, we are also able to reproduce the enhancement of neutron-proton 
pairs over neutron-neutron pairs due to the tensor interaction of the initial Hamiltonian (see right panel). Furthermore, in the 
present kinematic conditions there is factorization of the unitary transformation, which leads
to significant simplifications and an alternative interpretation of the universal high-momentum dependence and scaling~\cite{EricOperatorEvolv}.
This also illustrates that physical interpretations and intuition in general depend on the resolution, whereas the RG 
is the tool that allows us to connect the different pictures.

\begin{theacknowledgments}
I would like to thank R.\ J.\ Furnstahl for helpful comments on the manuscript. This work was supported by the NSF under 
Grant No.~PHY--1002478.
\end{theacknowledgments}

\bibliographystyle{aipproc}   

\end{document}